\documentstyle[preprint,eqsecnum,aps]{revtex}

\begin{document}
\draft

\title{Bose-Einstein condensation  in harmonic double wells}

\author{P. Capuzzi and E. S. Hern\'andez$^1$}
\address{Departamento de F\'{\i}sica, Facultad de Ciencias
Exactas y Naturales, \\
Universidad de Buenos Aires, RA-1428 Buenos Aires, \\
$^1$ and Consejo Nacional de Investigaciones Cient\'{\i}ficas y
T\'ecnicas, Argentina}

\maketitle

\begin{abstract}

	We  discuss Bose-Einstein condensation in harmonic traps 
where the confinement has undergone a splitting along one
direction. We mostly consider the 3D potentials consisting of
two cylindrical wells separated a distance 2$a$ along the
$z$-axis. For ideal gases, the thermodynamics of the confined
bosons has been investigated performing exact numerical
summations  to describe the major details of the transition and
comparing the results with the semiclassical  density-of-states
approximation. We find that  for large particle number and
increasing well separation, the condensation temperature evolves
from the thermodynamic limit value T$_c^{(0)}(N)$ to
T$_c^{(0)}(N/2)$. The effects of adding a repulsive interaction
between atoms has been examined resorting to the
Gross-Pitaevskii-Popov procedure and it is found that the shift
of the condensation temperature  exhibits different signs
according to the separation between wells. In particular, for
sufficiently large splitting, the trend opposes the well known
results for harmonic traps, since the critical temperature
appears to increase with growing repulsion strength.
\end{abstract}
\vspace{0.5truecm}
Pacs{\,\,03.75.Fi,05.30.Jp,32.80.Pj}
\narrowtext

\section{Introduction}

The recent  observation of Bose-Einstein condensation (BEC) of
atomic alkali gases subjected to  magnetic or magnetooptic
traps\cite{Rb,Na,Li} has triggered an important amount of
related work.  In fact, since the possibility of experimental
realization of BEC was put forward, sizeable theoretical effort
addressed both the quantum and the thermodynamical aspects of
this phenomenon when confined boson systems are involved.
Starting from the earliest approaches to this field
\cite{Groot,Rehr,Bagna1,Bagna2}, extensive  work has been devoted to the
various properties of harmonic traps 
\cite{BP,KT1,KT2,GH1,GH2,Kete1,Mul,HHR,HHA,KT3,Kete2,DPS1,DPS2,GPS1,Fetter},
which can account for most of the characteristic scales of
experimentally achieved BEC according to very intuitive scaling
arguments based on energy balance\cite{BP}. The main features of
BEC of ideal gases in either isotropic or anisotropic quadratic
potential wells have been analyzed by various authors
\cite{KT1,KT2,GH1,GH2,Kete1,Mul,HHR,HHA,KT3,Kete2},
paying special attention to the modifications of  the
transition patterns due to the  finite number of
trapped atoms.

The nature of the transition itself has been the subject
of some controversy. Kirsten and Toms have claimed that no
spontaneous symmetry breaking leading to BEC can happen in
confined potentials\cite{KT1,KT2}, as evidenced by the fact that
the chemical potential of the bosons does not reach the value of
the ground state energy at a finite temperature. However, this
behavior of the chemical potential does not prevent the
existence of a noticeable peak in the specific heat of the gas
at a well defined temperature, for dimensions other than unity,
coinciding with a sudden important rise in the ground state
occupation number.  This characteristic has been recognized by
various authors \cite{KT2,Pa,GH1} as the signature of a phase
transition. Furthermore, calculation methods based on exact
summations\cite{GH1,GH2,Kete1,Mul} and Euler-MacLaurin
approximation formulae\cite{HHR,HHA} yield results practically
identical  to those arising from more sophisticated treatments
\cite{KT1,KT2,KT3}.  
  Bose condensed gases in anisotropic traps have also deserved
attention, especially in view of the fact that experimental
devices lead to the possibility that the  definite stages of
condensation are governed by the most binding 1D
forces\cite{HHA,Kete2}. In particular, it has been clearly established that in
highly anisotropic potentials, the peak in the specific heat is
linked to the freezing of those degrees of freedom lying higher
in the energy spectrum, and macroscopic occupation of the ground
state of the system occurs at a lower temperature\cite{Kete2}.

The  effects of two particle interactions represented by a
single parameter, the scattering length, have been
examined\cite{Bagna1,DPS1,DPS2,GPS1,Fetter};  the solutions
of the Gross-Pitaevskii (GP) equation provide information upon
the condensate wave function,  and it has been shown that the
transition temperature is sensitive to the sign of the
interparticle forces in a way that opposes the expected results
for the free homogeneous gas\cite{GPS1}. The variational
solution to the GP equation\cite{Fetter} also provides an
interesting means to approach the various features  of BEC of
weakly interacting gases.

	More recently, double Bose condensates have become an
attractive field of research in view of the close resemblance
between these systems and other bistable devices familiar in
quantum optics\cite{Sci97}. The properties of double traps have
been mostly investigated from the perspective of these
analogies, keeping in mind that due to the large separations
between the experimentally split traps and the
significant height of the halving barriers, the condensate wave
function essentially corresponds to two nonoverlapping wells.
It is our present purpose to perform a detailed study of the quantum and
thermodynamic aspects of these bistable traps in terms of the
shape of the barrier and the well separation in the framework of a
simple description.  For this sake, in Sec. 2 we present the
essential formulae for the quantum mechanics and the
thermodynamics of an ideal gas in a harmonic double well, while
the results of the calculation are presented in Sec. 3. Section
4 is devoted to the weakly interacting gas described by means of
the GP + Popov formalism.    The conclusions are
summarized in Sec.  4.

\section{The harmonic double well}

	The spectrum of the 1D harmonic double well
with potential
\begin{equation}
V(z)= \frac{m\,\omega^2}{2}\,\left(\vert z \vert- a \right)^2
\label{Vz}
\end{equation}
can be  represented\cite{merz} by the parabolic cylinder functions
$D_{\nu}[\sqrt{2 m\,\omega\,/\hbar}\,(|z|-a)]$
with eigenvalues $\varepsilon_{\nu}= \hbar\,\omega(\nu+1/2)$.
The quantum numbers $\nu$ are determined establishing that the
eigenfunctions be either  even or odd and it is found that the
spectrum evolves from a purely harmonic one at separation $a$
= 0 to a doubly degenerate  harmonic one of the same frequency ,
as $a$ grows indefinitely. Furthermore, we shall consider 2D
({\it i.e.}, $\omega_y = 0$) and 3D traps of the form
\begin{equation}
V({\bf r}) = \frac{m}{2}\,\left[\omega_x^2\,x^2+\omega_y^2\,y^2+
\omega_z^2\,\left(\vert z \vert - a \right)^2\right]
\label{potencial}
\end{equation}
A numerical computation of the energy levels of (\ref{Vz})
enables us to examine the thermodynamic properties of a confined
boson gas  by straightforward summation of the relation defining
the internal energy,
\begin{equation}
U=\sum_i\,N_i\,\varepsilon_i
\label{Uex}
\end{equation}
where $N_i= \left[ \displaystyle e^{\beta (\varepsilon_i -
\mu)} - 1\right]^{-1}$ is the boson occupation number at
temperature $T= 1 /\beta$ for the level with energy $\varepsilon_i = \hbar
\,(n_x\,\omega_x +n_y\,\omega_y+\nu\,\omega_z+3/2)$, once the
particle number equation
\begin{equation}
N = \sum_i\,n_i
\label{Nex}
\end{equation}
has been solved to determine  the chemical potential
$\mu$ for the given value of $N$.  The specific heat is
afterwards computed  differentiating the energy with respect to
temperature.

In this work we compare the results of the exact numerical
calculation with those of the semiclassical method derived by
Bagnato {\it et al}\cite{Bagna1}, which in the present case
consists of replacing the summation in (\ref{Nex}) by
\begin{equation}
N=N_0 + N_1+\sum_j\,z^j\,\int d \varepsilon\,\rho(\varepsilon)\,
e^{-\frac{\displaystyle j\,\varepsilon}{\displaystyle k\,T}}
\label{Nbagna}
\end{equation}
where $\rho(\varepsilon)$ is the classical extension of the
density of states for single particles in the well. Notice that
in Eq. (\ref{Nbagna}) we explicitely separate the populations
$N_0$ of the ground state and $N_1$ of the first excited state,
since as the height of the barrier $V_0 = m\,\omega^2\,a^2/2$
increases, these two energy levels approach the doubly
degenerate ground state of the infinitely distant wells.  The
density of states is computed according to the prescription
given in Ref.
\cite{Bagna1}, which for an arbitrary confining D-dimensional
potential $V({\bf r})$ can be cast into the most general form
\begin{equation}
\rho_D(\varepsilon)=\frac{C_D}{h^D}\,2^{D/2-1}\,m^{D/2}\,
\int d^D r \left[\varepsilon - V({\bf r})\right]^{D/2-1}
\label{rhoD}
\end{equation}
where the integral is to be carried within the classically allowed
region. In view of the relation
\begin{equation}
\rho_D(\varepsilon)=\int_0^{\varepsilon} d
\varepsilon'\,\rho_{D-1}(\varepsilon-\varepsilon')\,\rho_1(\varepsilon')
\label{convo}
\end{equation}
we only need to compute the 1D density for the double well. 
The  3D case is  studied convoluting $\rho_1$ with
\begin{equation}
\rho_2(\varepsilon)=\frac{1}{\hbar\,\omega}\,\left(\frac{\varepsilon}
{\hbar\,\omega}+ 1\right)
\label{rhobuena}
\end{equation}
for $\omega=\omega_x=\omega_y$, which   yields a formula for the
number of particles in two dimensions that coincides with the
high temperature limit of the exact summation (see Appendix). It
should be noticed that  the semiclassical prescription of Ref.
\cite{Bagna1} only gives the first term of Eq. (\ref{rhobuena})
for the 2D density.

\section{The symmetric double well}

	We have investigated BEC of an ideal boson gas confined
in the  trap described in the preceding section for dimensions
D=1 to 3, comparing the results of the exact summations with
those of the approximate density-of-states method, in terms of
the well splitting $a$ and the number of particles $N$. In the
3D case, we also examine the anisotropic potentials mostly
related to experimental situations; notice that in spite of the
symmetry breaking along the $z$-direction, we shall consider the
potential to be isotropic when $\omega_x = \omega_y = \omega_z$
and anisotropic otherwise. The analytical expressions for the
semiclassical densities of states computed according to the
prescription (\ref{rhoD}) and the corresponding numbers of
particles are summarized in the Appendix.

The specific calculations indicate that the 1D well exhibits the
characteristic sharp increase of the occupation number $N_0$
around a temperature T$_c$, however this is the usual
seudotransition that is not accompanied by a peak in the
specific heat, as shown in Fig. 1 for various separations
between harmonic wells (hereafter indicated by the dimensionless
variable $z_0= \sqrt{2 m\,\omega/\hbar}\,a$). We can appreciate
the moderate sensitivity of this quantity to the splitting
$z_0$.  We also compare the exact ground state
$\psi(z)$ of the double well with the wave function
\begin{equation}
\tilde{\psi}(z) = {\cal N}\,\left[\psi_0(z-a) + \psi_0(z+a)\right]
\label{psi0}
\end{equation}
where $\psi_0( z \pm a)$ represents the ground state of
a single harmonic well of the same frequency centered at $\pm a$
and ${\cal N}$ is a normalization factor. This is illustrated in
Fig. 2, where we plot, for $z_0=$1 and 3, (left and right
columns, respectively) the functions $\psi(z)$ and
$\tilde{\psi}(z)$ in full and dashed lines, respectively.  In
the upper plots, each single ground state wave function
$\psi_0(z \pm a)$ has been individually normalized, whereas in the
lower ones the sum (\ref{psi0}) has been normalized to unity.
The dotted lines  indicate each separate contribution to
$\tilde{\psi}(z)$. It is clear that for every well separation
here considered, the exact wave function resembles almost
exactly the normalized combination (\ref{psi0}) shown in the
lower pictures.

Let us now  consider a 3D isotropic trap; no substantial
qualitative differences have been observed in the 2D case. In
Fig. 3 we exhibit the occupation numbers $N_0$ and $N_1$ as
functions of the temperature in units of the critical
temperature $T_c^{(0)}$ corresponding to infinite number of
atoms, for well separations $z_0=0$ ({\it i.e.}, a single
harmonic well) and 1, as well as for total particle numbers
N=100 and 10000.  We  observe a rather sharp transition  at a
temperature slightly lower than unity for sufficiently large
number of particles; furthermore, this transition temperature
decreases with growing barrier height.  On the other hand, the
larger the value of $z_0$, the closer the resemblance between
$N_0$ and $N_1$ even for temperatures substantially below
$T_c^{(0)}$; this fact reflects the quasidegeneracy of the
ground state and the first excited state for large finite values
of $z_0$, which evolves into complete degeneracy at infinite
separation. This can be visualized in Fig. 4, where we have
selected N=1000 and  various barrier heights $V_0$, which allow
us to appreciate the significant decrease of the transition
temperature that occurs as the wells split apart.

	The existence of a transition becomes evident as one
analyzes the specific heat. Figs. 5 and 6 correspond to the same
parameters as Figs. 3 and 4, respectively; while lines indicate
numerical results, symbols locate the calculations performed
with the semiclassical expressions listed in the Appendix. We
observe that the agreement between exact and approximate results
is almost perfect at the lowest temperatures, whereas for high
temperatures, that agreement holds only for large particle
numbers. For such numbers the semiclassical approach
locates the transition to an acceptable accuracy, except for
some sligth overestimation of both the characteristic
temperature and the height of the peak; however, this departure
is very noticeable for small quantities of atoms.  In Fig. 6 we
may notice the consequences of increasing the well separation:
the major effect takes place for a splitting $z_0 =3$, with only
minor differences as one enlarges  this number.

We have also performed calculations for anisotropic traps; as an
illustration, in Fig. 7 we show the specific heat for 1000 atoms
and  well separations $z_0$ = 0 and 1.
 The various curves correspond to the
aspect ratios of the JILA\cite{Rb} and MIT traps\cite{Na},
namely $\omega_z/\omega_{x,y}= \sqrt{8}$ and ($\omega_z/\omega_x=3.2$, 
$\omega_z/\omega_y=1.8$), respectively, and
to an isotropic well with a frequency $\omega = 0.56\,\omega_z$ equal to
the MIT geometric mean.  We realize that the larger the
anisotropy, the smaller is the transition temperature and the
lower is the height of the peak. This also happens when one
switches from the single to the double well, however to a less
significant extent as compared to the anisotropy effect.

\section{The weakly interacting gas in the double well}

	The influence of weak to moderate interactions between
trapped particles can be investigated in a GP mean field
approach generalized for finite temperatures. Within this
formalism, the condensate wavefunction $\psi(r)$ is obtained
from the  nonlinear equation
\cite{griffin}

\begin{equation}
\hat{h}_0\,\psi(r)=\mu\,\psi(r)
\label{GPE} 
\end{equation}
with
\begin{equation}
\hat{h}_0=\left\{ -\frac{\displaystyle \nabla^2}{2\,m} +V(r)+g\,[n_c(r)+2\,
\tilde{n}(r)]\right\}
\end{equation}
where we have made the usual decomposition of the density
into condensate and noncondensate contributions, i.e.,
$n(r)=n_c(r)+\tilde{n}(r)$, and $n_c(r)=|\psi(r)|^2$.  The terms
involving the interaction strength $g$ arise from  a two-body
seudopotential $g\,\delta({\bf r})$.  In the $s$-wave
approximation, which is adequate for very dilute gases, one has 
$g=4\pi\,\alpha\,\hbar^2/\!m$, being $\alpha$ the scattering
length.  

Within the Popov approximation of the Hartree-Fock-Bogoliubov (HFB) theory, 
the  excitations of a dilute system of bosons at low and intermediate
temperatures
are given by the coupled eigenvalue equations 
\begin{eqnarray}
\hat{\cal L}\,u_i(r)-g\,n_c(r)\,v_i(r) &=& E_i\,u_i(r)\nonumber \\
\hat{\cal L}\,v_i(r)-g\,n_c(r)\,u_i(r) &=&-E_i\,v_i(r) 
\label{HFB}
\end{eqnarray}
which define the quasiparticle  energies $E_i$ 
and amplitudes $u_i$ and $v_i$, with
\begin{equation}
\hat{\cal L}\equiv\hat{h}_0+g\,n_c(r)
\end{equation}
 The noncondensate density is related to the quasiparticle properties
according to \cite{griffin,fetter2}
\begin{equation}
\tilde{n}(r)=\sum_i\,\left\{|v_i({\bf r})|^2 + \left[|u_i({\bf r})|^2 +
 |v_i({\bf r})
|^2\right]\,N(E_i)\right\}
\label{noncond}
\end{equation}
where $N(E_i)=\left(e^{\beta E_i}-1\right)^{-1}$ is the
Bose-Einstein distribution of the quasiparticles.

One must solve the coupled  GP  (\ref{GPE}) and HFB  (\ref{HFB})
equations, using the self-consistent densities $n_c(r)$ and
$\tilde{n}(r)$ for  fixed number of particles  $N$. For this
sake, we use the decoupling procedure introduced by Hutchinson
{\it et al} \cite{Hutch}, and expand $\tilde{h}_0$ in a basis of
eigenfunctions of the 2D harmonic oscillator on the $(x,y)$
plane times a discretized basis set for the 1D double well along
the $z$-axis.  To illustrate this procedure, in Fig. 8 we plot
the condensate fraction as a function of the temperature, for
100 particles at various barrier heights and positive
interaction strengths $s=\alpha\,\sqrt{m\,\omega_z/\hbar}$.  We
observe that in all cases, the system reaches BEC at
temperatures below the transition temperature for a single well
with no interactions.  However, depending on the value of $z_0$
two different behaviors occur. When $z_0$ is small, the
condensation temperature decreases, the stronger the
interaction, in agreement with  the well known result for the
harmonic trap\cite{Bagna1,GPS1}.  Conversely, if $z_0$ is high
enough, an increase of $s$ yields a higher critical temperature.
This feature can be interpreted if we consider  the  single
particle spectrum in the mean field, whose ground and first
excited states are depicted in Fig.  9 for a temperature T=0.26
T$_c^{(0)}$ and N=100. The overall trend is similar at any
temperature and this choice is a convenient one in view of the
displayed scale.  We realize that these levels also exhibit two
tendencies:  for small well separations, as the interaction is
enlarged the energy levels get closer, giving a  more degenerate
spectrum and a higher density of states, which in turn demand
further cooling in order to achieve BEC. On the other hand, if
$z_0$ is sufficiently large (above roughly $z_0$=2.5 in Fig. 9)
the ground and first excited states of a double well are already
quasi-degenerate; in this case, the effect of the interaction as
indicated in this figure is to uniformly raise the whole spectrum.
As a consequence, due to level crossing, particles occupying the
first excited state of the free gas can be promoted to the
interacting ground state at no thermodynamic cost, giving thus
rise to a moderate increase of the condensation temperature.

\section{Discussion and summary}

	In this work we have discussed  the characteristics of
Bose-Einstein condensation in harmonic wells where the
confinement has undergone some splitting along one of the three
directions. We mostly consider the 3D potentials consisting of
two cylindrical wells separated a distance 2$a$ along the
$z$-axis. For ideal gases, the thermodynamics of the confined
bosons has been investigated performing exact numerical
summations for the particle number, which enables us to
determine the chemical potential, and for the total energy,
after which we derive the specific heat by means of a numerical
differentiation  with respect to temperature.  This procedure
allows us to describe the major details of the transition and to
compare the results with the approximation corresponding to 
replacement of exact summations by integrals  weighted by a
semiclassical density of states.

	The general results can be summarized as follows. We
find that the semiclassical approach is capable of locating the
transition, {\it i.e.}, the peak in the specific heat, with
higher accuracy the larger the amount of trapped atoms. The
overestimation in both the precise value of the transition
temperature and  the height of the peak is rather pronounced for
small systems and becomes progressively less important as the
system approaches the thermodynamic limit.  It becomes clear
that the effect of halving the population  is to lower the
transition temperature; for large particle number, the trend of
this temperature as the well separation grows from
zero to infinity is clearly to sweep the path between critical
temperatures $T_c(N)$ and $T_c(N/2)$. This general behavior is
independent of the dimensionality of the system; however, just
as in the single well problem \cite{GH1,Kete1}, no strict phase
transition takes place in the 1D potential. In fact, in this
case one finds that in spite of the important rise in the slope
of the occupation number $N_0$ observed at a rather well defined
temperature,  the specific heat increases monotonically 
 towards the classical limit.

	The effects of adding a repulsive interaction between
atoms has been examined resorting to the GP + Popov procedure
which yields the condensate and noncondensate densities together
with the single particle spectrum. It is found that the shift of
the condensation temperature  exhibits different signs according
to the size of the separation between wells on the $z$ axis; in
particular, for sufficiently large splitting, the trend opposes
the well known results for  harmonic traps, since the critical
temperature appears to increase with growing repulsion strength.
This unexpected feature can be explained examining   the
spectrum of the atoms in the mean field as a function of the
well separation. One can realize that in the interacting system,
not only the position of the energy levels, but the  size of the
gap between the ground and the first excited state, change
substantially when the wells split apart. The level crossing
that occurs for  large separations causes atoms in the excited
state to move into the new ground state, with a consequent
increase in the condensation temperature.

An experimental realization of the double well is the MIT trap
\cite{Na} characterized by a large separation  between
wells and by a tunable barrier height. One might then
wonder to what an extent the  results here presented might
be sensitive to independent choices of the  location $\pm a$ of
the minima and the barrier magnitude $V_0$, and intend to
develop more detailed models for the specific MIT double trap.
Among the simplest choices, one could quote  i) a combination of
two symmetric harmonic wells smoothly joined  at $z = \pm a_m$ by a
quadratic barrier and ii) the double well of the previous
sections deformed by a central constant cutoff  with value $V_0$
for $|z| \le a_m$.  We have performed calculations similar to those
presented in section 3 for the above potentials, employing
standard magnitudes of the original MIT trap \cite{Na}. We have
found that the thermodynamics of the ideal gas remains identical
to the previously analized double well case, in other words, the
transition temperature and the overall shape of the occupation
numbers and specific heat (Figs. 3 to 6) are not sensitive to
the decoupling between the location of the minima  and the
barrier size, within the range of values of interest.

	To summarize, we remark that the statistical mechanics
of a bistable  system  can be examined on identical grounds as
the case of a single equilibrium state, with an additional
parameter, namely the separation between wells,  as the agent of
the evolution from a given number of trapped atoms up to the
point where  one exactly halves the population in the infinite
separation limit.

\acknowledgements

This paper was performed under grants PICT 0155 from Agencia
Nacional de Promoci\'on Cient\'{\i}fica y Tecnol\'ogica of Argentina
and EX/0100 from Universidad de Buenos Aires.

\section*{appendix}

We collect here various useful formulae concerning the
thermodynamics of ideal gases in harmonic traps. We  display
here  the semiclassical results obtained from the
density-of-states method of Ref. \cite{Bagna1} (Eq.
(\ref{Nbagna}) for the isotropic D-dimensional oscillator,
\vspace{1pc}

\noindent{\it Particle number:}
\begin{equation}
N=N_0+\left(\frac{k\, T}{\hbar\,\omega}\right)^D\,g_{ D}(z)
\end{equation}

\noindent{\it Internal energy:}
\begin{equation}
 U=D\,\hbar\,\omega\,\left(\frac{k
T}{\hbar\,\omega}\right)^{D+1}\,g_{ D+1}(z)
\end{equation}

\noindent{\it Specific heat:}
\begin{equation}
\frac{C}{k}=D\,\left(\frac{k \,T}{\hbar\,\omega}\right)^D\,\left[(D+1)\,
g_{ D+1}(z)-D \frac{g_{ D}^2(z)}{g_{ D-1}}(z)\right]
\end{equation}
whereas the exact summations give, in the high temperature limit
$\hbar\,\omega << k \,T$
\widetext
\begin{equation}
N=N_0+\left(\frac{k \,T}{\hbar\,\omega}\right)^D\,\sum_{n=1}^{D}\,
(-)^{n+1}\,g_{ n}\left[z\,e^{\displaystyle -(n-D/2)\,\hbar\,\omega/k\,T}\right]
\end{equation}
\narrowtext
In this limit one can take advantage of the property of the Bose special
functions $g_i(z)$
\begin{equation}
g_i\left(e^{\displaystyle \frac{\mu+\delta \mu}{k\,T}}\right) \approx
g_i(z)+ \frac{\delta \,\mu}{k\,T}\,\frac{g_{i-1}(z)}{z}
\end{equation}
for $D \ge 2$ and write
\begin{equation}
N=N_0+\left(\frac{k\,T}{\hbar\,\omega}\right)^D\,\left[g_D(z) +
\frac{D}{2}\,\frac{k\,T}{\hbar\,\omega}\,\frac{g_{D-1}(z)}{z}\right]
\label{Naltat}
\end{equation}
from where the transition temperature is obtained specifying
$N_0$ = 0 together with $z = 1$. 

	For the isotropic potential with the double well on the
$z$ direction, we have the following cases
\vspace{1pc}

\noindent{\it Dimension D = 1:}

	The density of states takes the form
\widetext
\begin{equation}
\rho_{1D}(\varepsilon)=\left\{ 
\begin{array}{ll}
\frac{\displaystyle 2}{\displaystyle \hbar\omega}, & \varepsilon
< V_o \nonumber \\ & \nonumber \\
\frac{\displaystyle 1}{\displaystyle \hbar\omega} +
\frac{\displaystyle 2}{\displaystyle \pi}\,\frac{\displaystyle
1}{\displaystyle \hbar \omega}\,arcsin\left(
\sqrt{\frac{\displaystyle V_o}{\displaystyle \varepsilon}}\right), 
& \varepsilon \ge
V_o
\end{array}
\right.
\end{equation}
\narrowtext
and the number of particles is
\widetext
\begin{eqnarray}
N&=&N_0+N_1+\int_{\varepsilon _1^{+}}^\infty d\varepsilon\,\frac{ \rho
_{1D}(\varepsilon )}{e^{\beta (\varepsilon -\mu )}-1}
\nonumber
\\
 &=&\frac z{1-z}+\frac{z_1}{1-z_1}+\frac 1{\beta \hbar \omega }\,\ln \frac{%
1-z_m}{(1-z_1)^2}
\nonumber
 \\
&&+\frac 2\pi \,\frac 1{\hbar \omega }\,\int_{\varepsilon
_m}^\infty d \varepsilon\, arcsin\sqrt{%
\frac{V_o}{\varepsilon}}\,N(\varepsilon)
\label{N1D}
\end{eqnarray}
\narrowtext
\noindent with $z=\exp (\beta \mu ),
z_1 =\exp \left\{ \beta (\mu -\varepsilon _1)\right\},
z_m =\exp \left\{ \beta (\mu -\varepsilon _m)\right\}$ and  
$\varepsilon _m =\max \{\varepsilon _1,V_o\}$
The total energy can be expressed, similarly to (\ref{N1D}), as
a summation including an integral.  
\vspace{1pc}

\noindent{\it Dimension D=2:}

	We obtain
\widetext
\begin{equation}
\rho_{2D}(\varepsilon)=\left\{ 
\begin{array}{ll}
2\,\frac{\displaystyle \varepsilon}{\displaystyle (\hbar \omega)^2},
 & \varepsilon <
V_o \nonumber 
\\ 
 \nonumber \\
\frac{\displaystyle 1}{(\displaystyle \hbar \omega)^2} \left[ \varepsilon
+\varepsilon\,\frac{\displaystyle 2}{\displaystyle
\pi}\,arcsin(\sqrt{V_o/\varepsilon})+
\frac{\displaystyle 4}{\displaystyle
\pi}\,\sqrt{V_o}\,\sqrt{\varepsilon-V_o} \right], & \varepsilon \ge V_o
\end{array}
\right.
\end{equation}
\narrowtext
and
\widetext
\begin{eqnarray}
N &=&\frac z{1-z}+\frac{z_1}{1-z_1}+\frac 1{(\beta \hbar \omega )^2}\,\left[
2\, g_2(z_1)-g_2(z_m)\right]   \nonumber \\
&&+\frac 2\pi \,\frac 1{\beta (\hbar \omega
)^2}\,\int_{\varepsilon _m}^\infty d \varepsilon\,arcsin\left(
\sqrt{\frac{V_o}e}\right)\, \ln \left[ 1-\exp \left( -\beta (e-\mu
)\right) \right]
\end{eqnarray}
\narrowtext
\noindent where $g_n(z)$ is the usual Bose function\cite{Huang}
\vspace{1pc}

\noindent{\it Dimension D=3:}

	The corresponding expressions are
\widetext
\begin{equation}
\rho _{3D}(\varepsilon)=\left\{ 
\begin{array}{ll}
\frac{\displaystyle \varepsilon^2}{\displaystyle (\hbar \omega )^3}, &
\varepsilon<V_o\nonumber 
\\ 
\nonumber 
\\
\frac {\displaystyle 1}{\displaystyle 2(\hbar \omega
)^3}\,\left[ \varepsilon^2+\frac {\displaystyle 2}{\displaystyle \pi}\,
\varepsilon^2\,arcsin\left(\sqrt{\frac{\displaystyle V_o}
{\displaystyle \varepsilon}} \right)\right.  &
\nonumber 
\\
\nonumber
\\
 +\left. \frac {\displaystyle 2}{\displaystyle 3}\,
\frac{\displaystyle \sqrt{V_o}}{\displaystyle \pi}\,
\sqrt{\varepsilon-V_o}\,(5 \varepsilon-2V_o)\right], 
 & \varepsilon\ge V_o
\end{array}
\right.   \label{ro3d}
\end{equation}
\narrowtext
and
\widetext
\begin{eqnarray} N &=&\frac z{1-z}+\frac{z_1}{1-z_1}+\frac
1{(\beta \hbar \omega )^3}\,\left\{ 2\ g_3(z_1)-g_3(z_m)+\beta
\hbar \omega \left[ 2\ g_2(z_1)-g_2(z_m)\right]
\right\}   \nonumber \\
&&+\frac 2\pi\, \frac 1{\beta ^2(\hbar \omega
)^3}\,\int_{\varepsilon _m}^\infty d \varepsilon\,arcsin\left(
\sqrt{\frac{V_o}{\varepsilon}}\right)\, \left[ g_2(z_\varepsilon )+\beta \hbar
\omega \ g_1(z_\varepsilon )\right] 
\end{eqnarray}

\narrowtext

\section*{Figure Captions}
\begin{description}
\item[Figure 1.-] The specific heat of a 1D boson gas trapped in a 
double well (in units of the Boltzmann constant $k_B$) as a
function of the reduced temperature $T/T_c^{(0)}$ for different
separation between potential minima.
\item[Figure 2.-] The exact ground state probability $\vert
\psi(z) \vert^2$ (full lines) of particles in the double trap as
a function of the dimensionless coordinate $z/a$, for  well
separations $z_0$=1 (left column) and $z_0$=3 (right column).
Dashed lines correspond to the sum of two normalized single well wave
functions (upper plots) and the normalized sum (lower plots),
while  dotted lines indicate the individual ground states.
\item[Figure 3.-] Occupation numbers of the ground and first
excited state for total particle number N=100 (thin lines) and
10000 (thick lines) as functions of $T/T_c^{(0)}$ for the single
harmonic well and for a separation equal to unity.
\item[Figure 4.-] Occupation numbers of the ground and first
excited state for 1000 atoms and various barrier heights.
\item[Figure 5.-] Specific heat for the same situation depicted in Fig. 3.
Circles and dots indicate the results of the semiclassical approximation.
\item[Figure 6.-] Same as Fig. 5 for the conditions of Fig. 4.
\item[Figure  7.-] Specific heat of anisotropic traps for  the
aspect ratios of the JILA and MIT traps, compared to an
isotropic situation.
\item[Figure 8.-] Ground state occupation number of a 3D double well
including weak two-body interactions, computed in the GP+Popov
 approximation.
\item[Figure 9.-] Energies of the ground and first excited states (full and
dotted lines, respectively) of interacting particles in an isotropic double
well, as functions of the separation $z_0$, for various interaction
strengths from s=0 up to s=10$^{-2}$ (bottom to top) in steps
$\Delta s$=10$^{-3}$. The dashed line indicates the barrier height $V_0$. 
Energies are given in units of $\hbar \omega$.
\end{description}
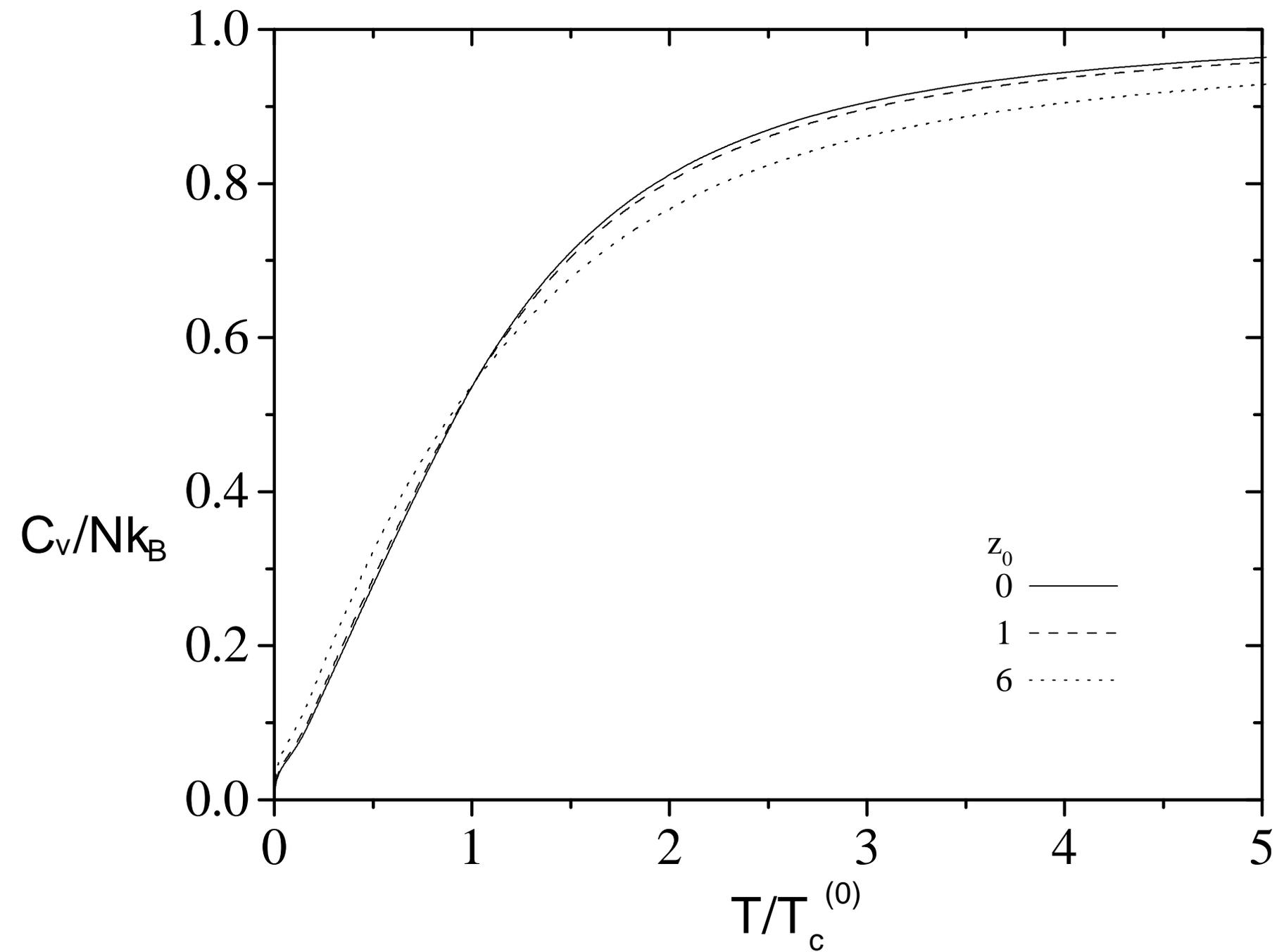
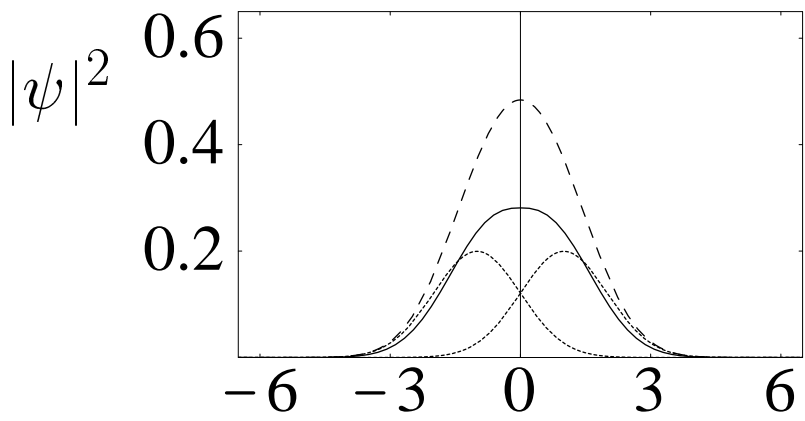
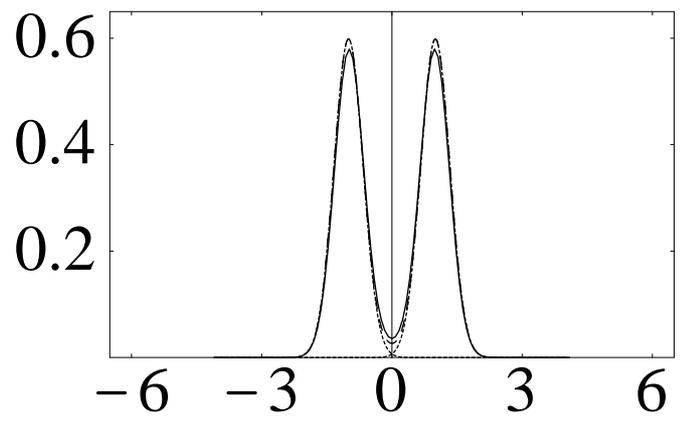
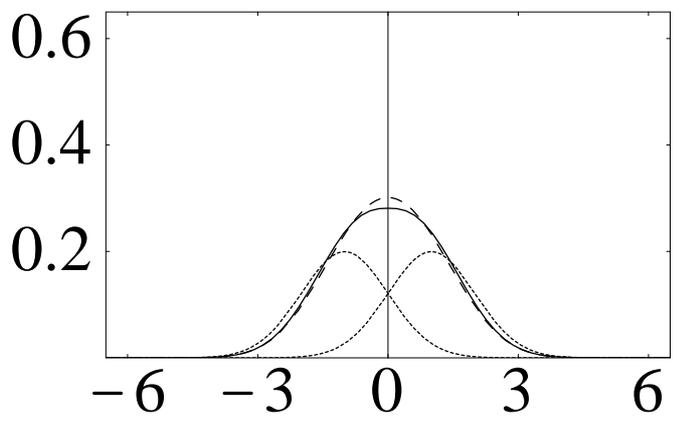
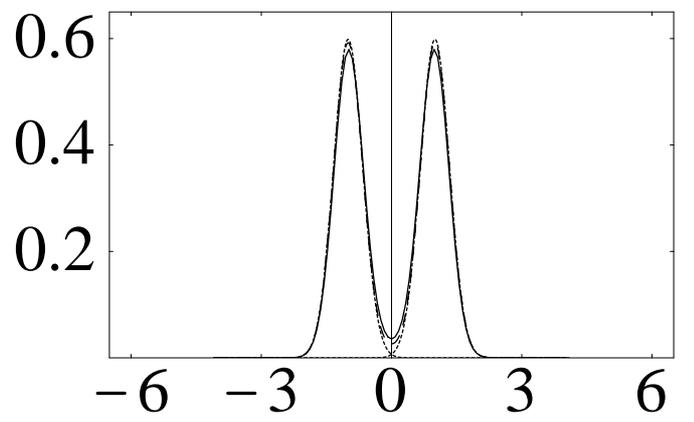
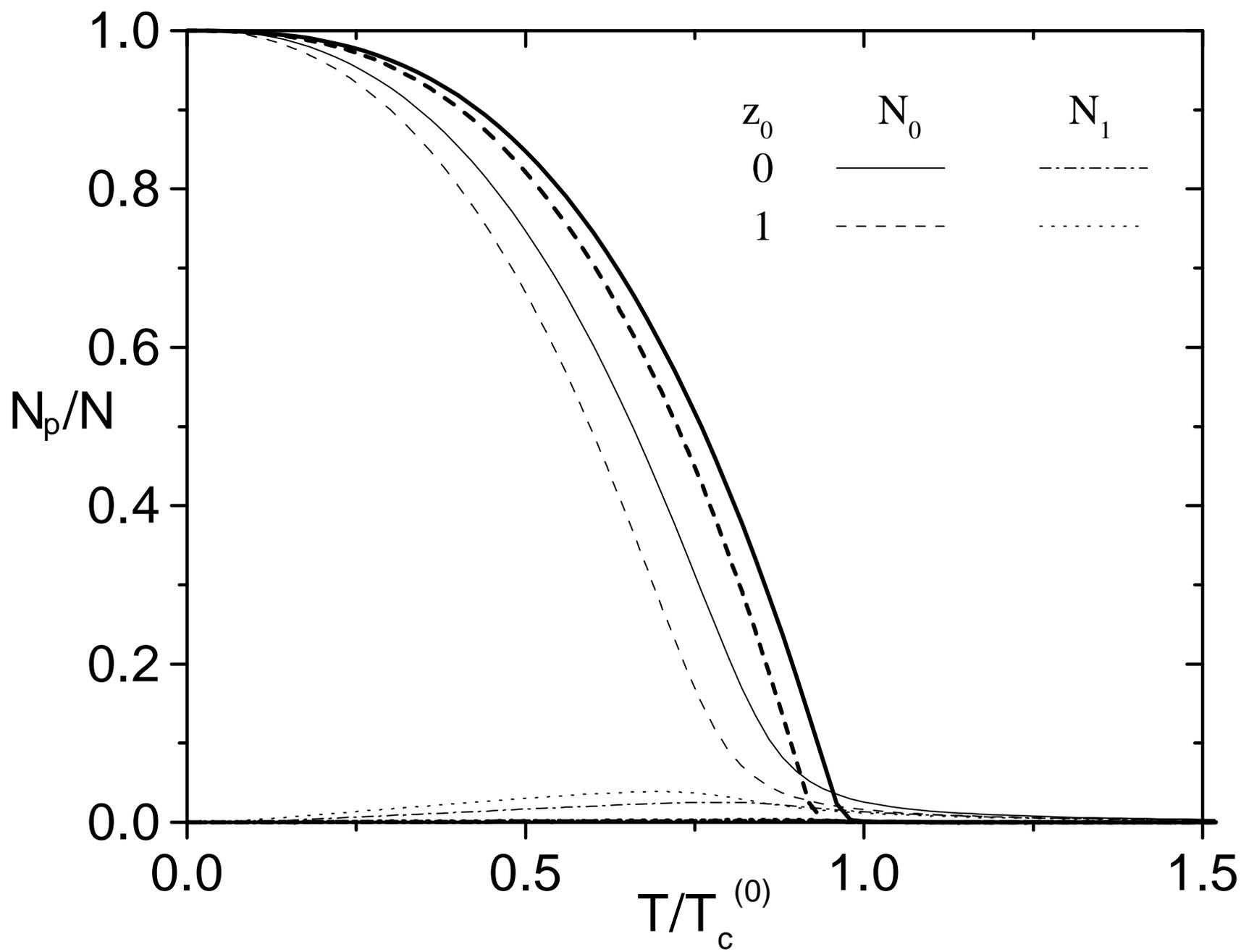
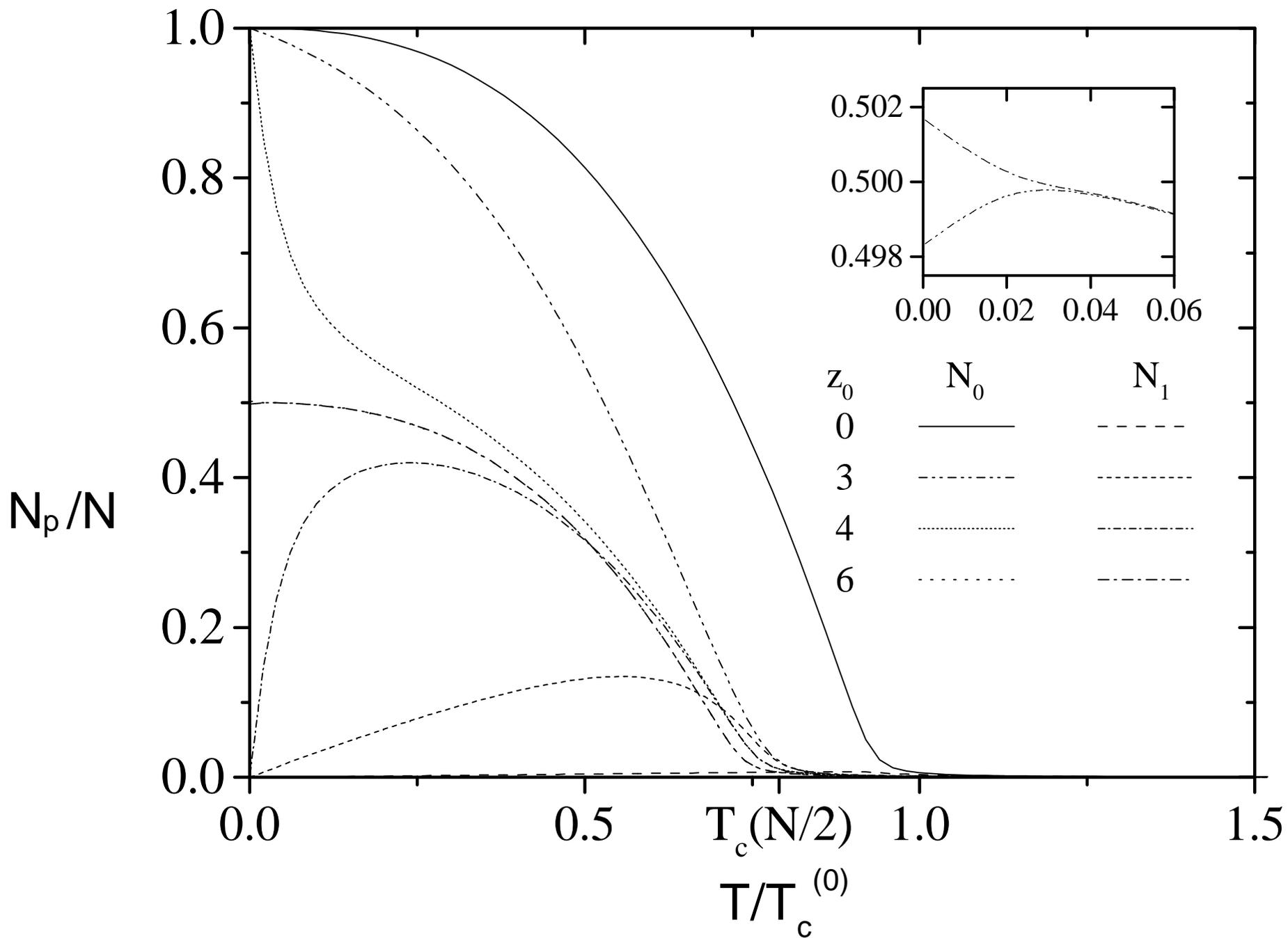
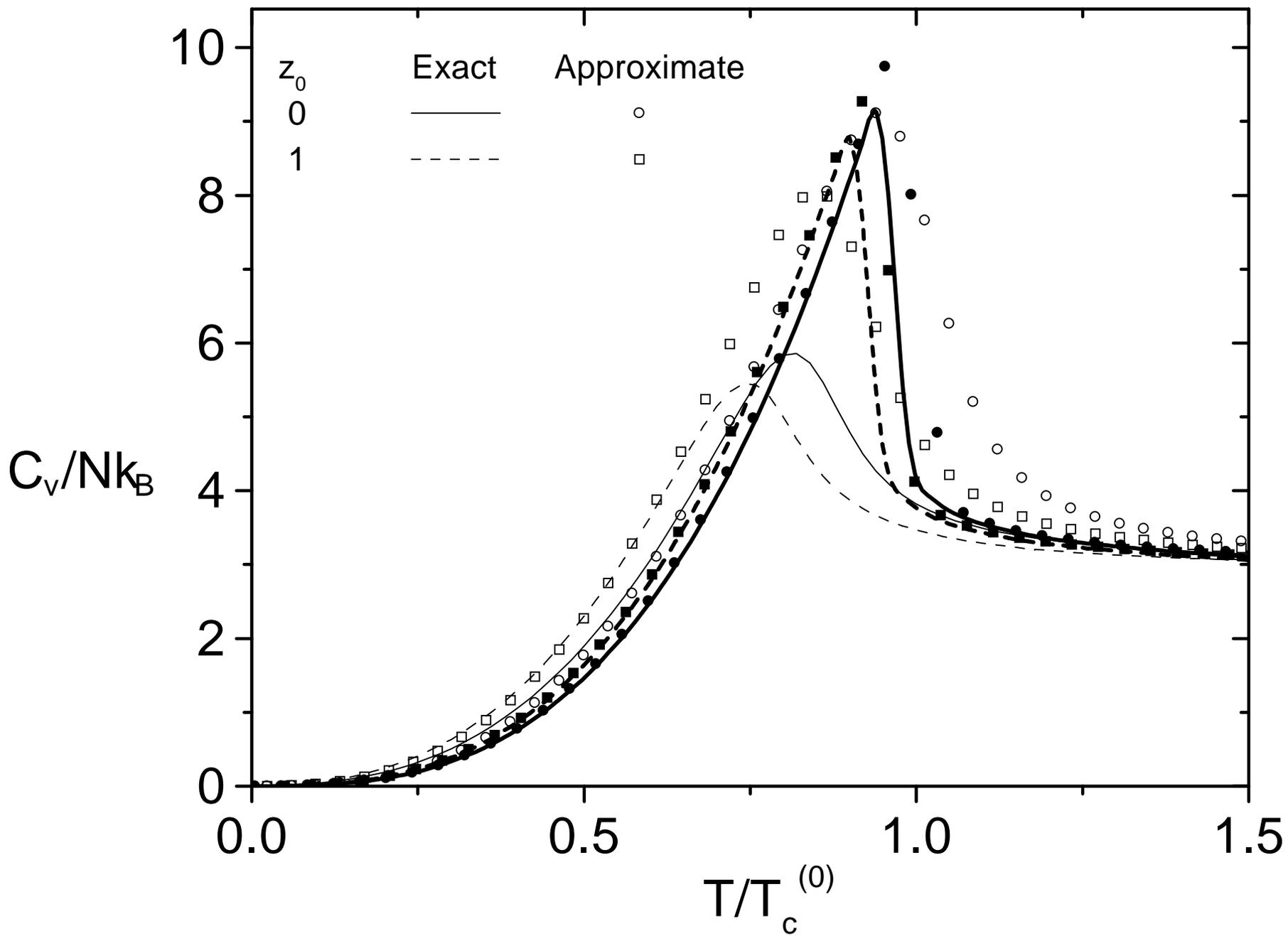
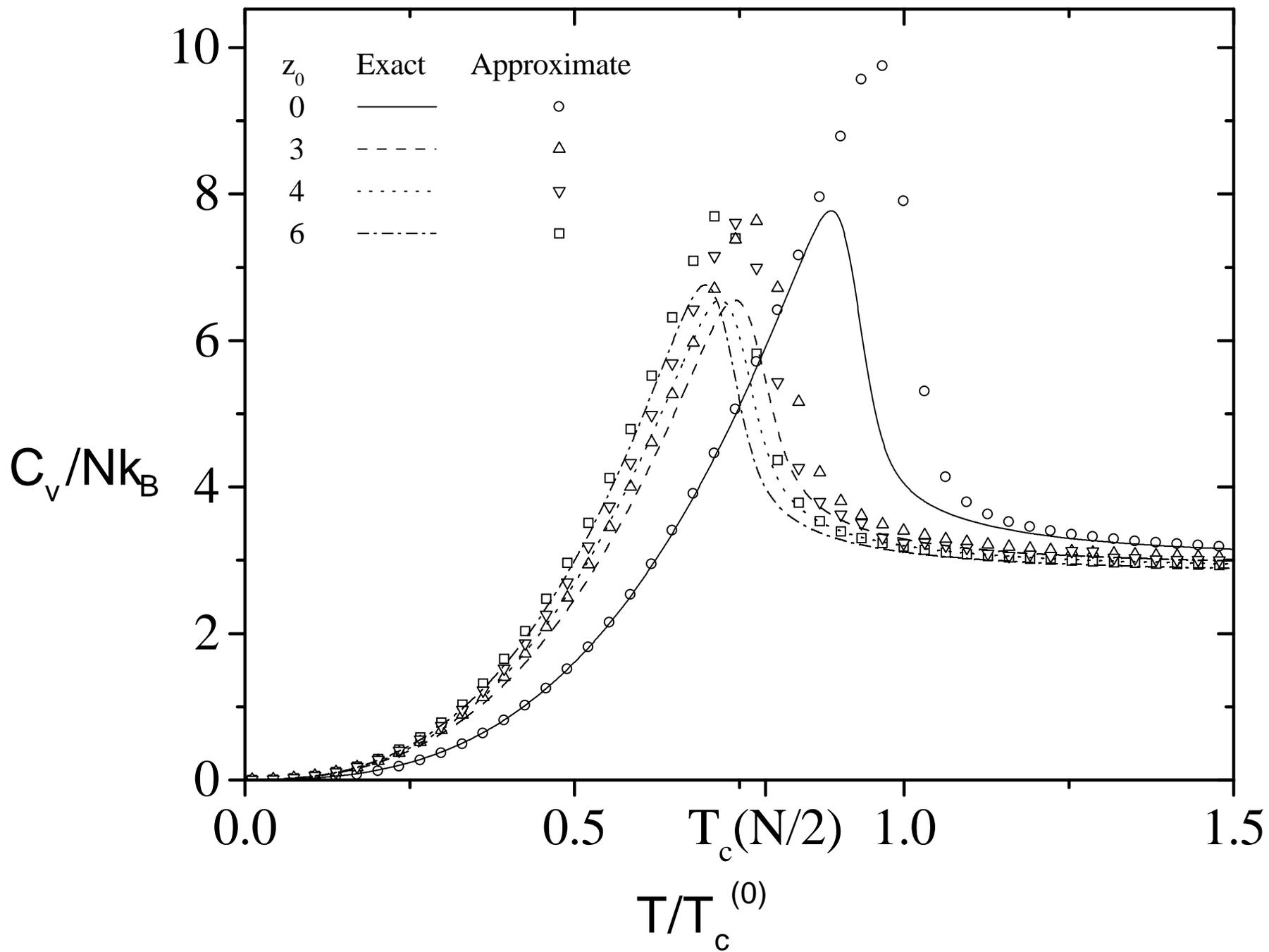
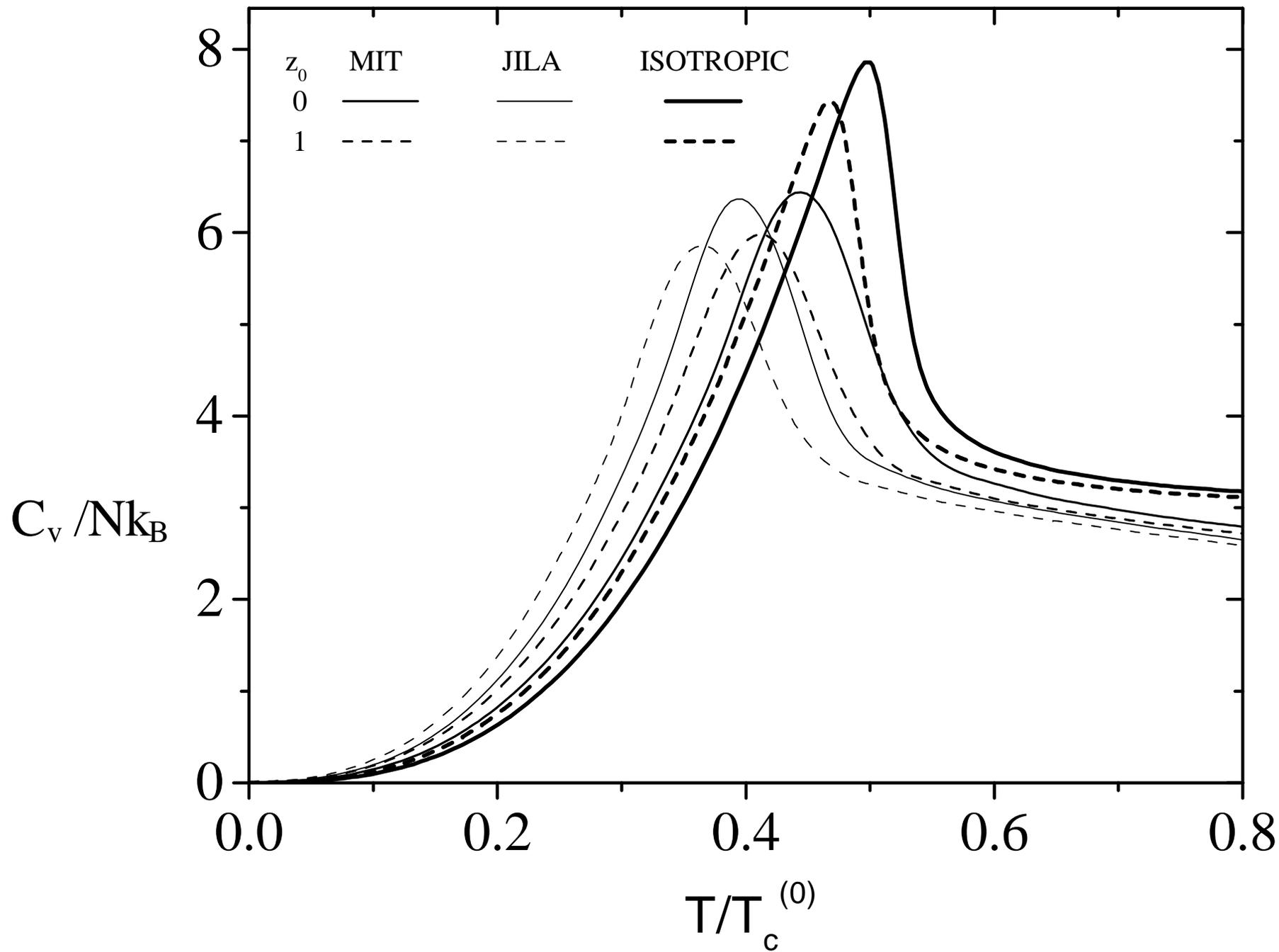
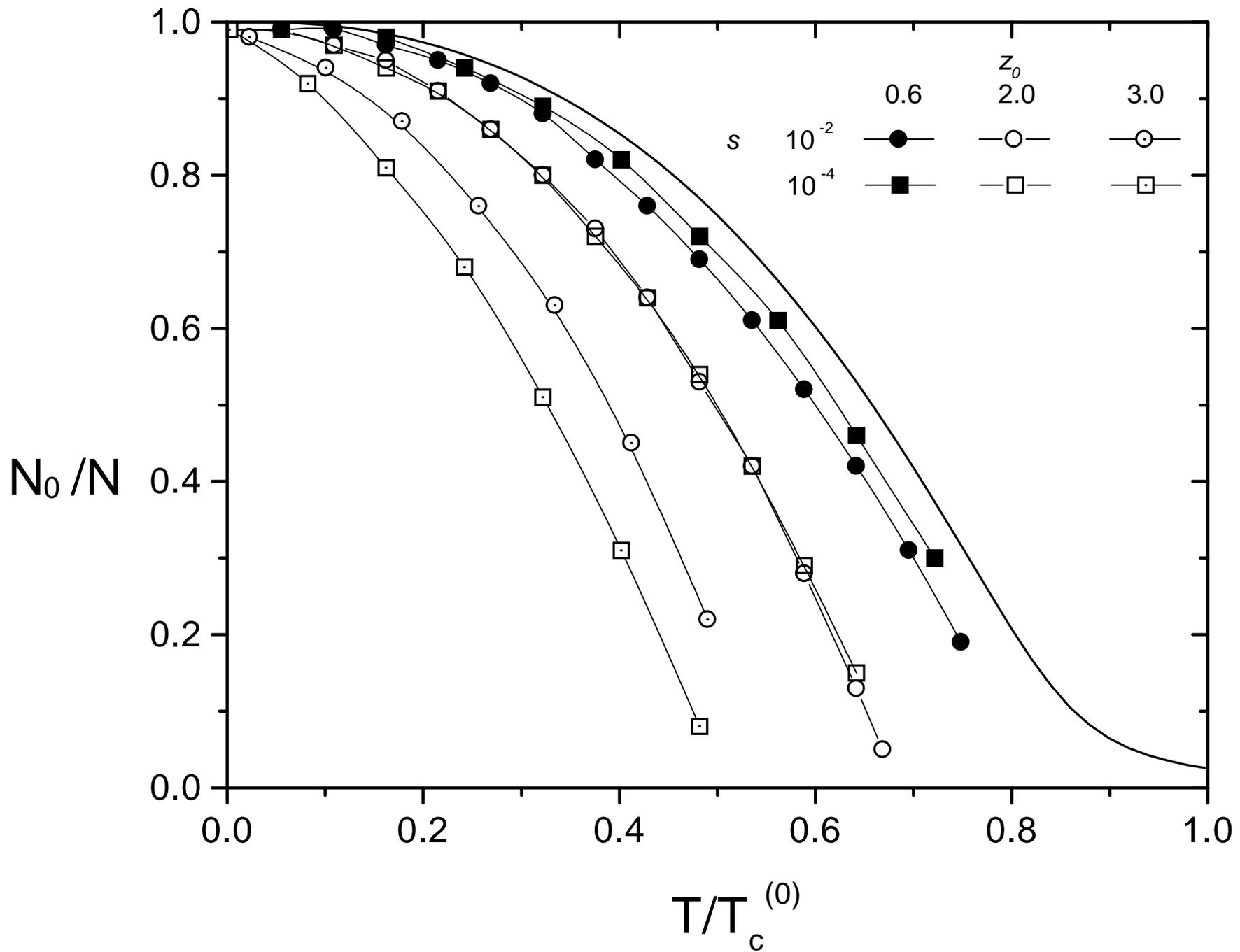
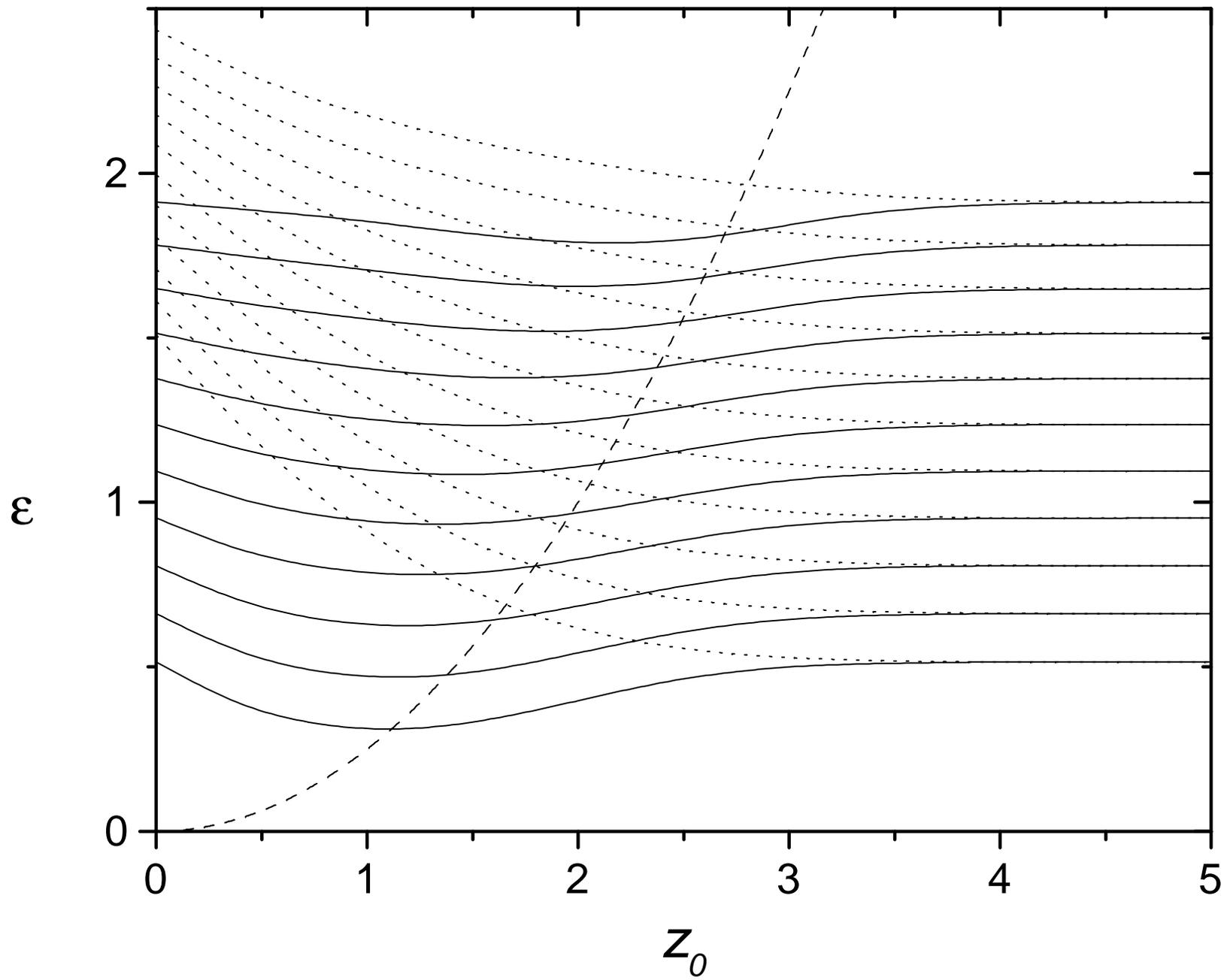
\end{document}